\documentclass[conference]{IEEEtran}
\IEEEoverridecommandlockouts

\usepackage{cite}
\usepackage{hyperref}
\usepackage{amsmath,amssymb,amsfonts}
\usepackage{algorithmic}
\usepackage{graphicx}
\usepackage{balance}
\usepackage{textcomp}
\usepackage[table]{xcolor} 
\usepackage{colortbl}  
\usepackage{longtable}
\usepackage{array}
\usepackage[most]{tcolorbox}
\usepackage{enumitem}

\def\BibTeX{{\rm B\kern-.05em{\sc i\kern-.025em b}\kern-.08em
    T\kern-.1667em\lower.7ex\hbox{E}\kern-.125emX}}

\tcbset{textmarker/.style={
        enhanced,
        parbox=false,boxrule=0mm,boxsep=0mm,arc=0mm,
        outer arc=0mm,left=5mm,right=3mm,top=7pt,bottom=7pt,
        toptitle=1mm,bottomtitle=1mm}}

\newtcolorbox{noteBox}{textmarker,
    borderline west={4pt}{0pt}{gray},
    colback=gray!10!white}
\newcommand{\note}[1]{\begin{noteBox} #1 \end{noteBox}}

\newcommand{\orcidlink}[1]{\textsuperscript{\href{https://orcid.org/#1}{\includegraphics[scale=0.2]{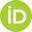}}}}
    
\begin{document}

\title{From Coders to Critics: Empowering Students through Peer Assessment in the Age of AI Copilots}

\author{\IEEEauthorblockN{Santiago Berrezueta-Guzman\orcidlink{0000-0001-5559-2056}}
\IEEEauthorblockA{\textit{Technical University of Munich}\\
Heilbronn, Germany \\
s.berrezueta@tum.de}
\and
\IEEEauthorblockN{Stephan Krusche\orcidlink{0000-0002-4552-644X}}
\IEEEauthorblockA{
\textit{Technical University of Munich}\\
Munich, Germany \\
krusche@tum.de}
\and
\IEEEauthorblockN{Stefan Wagner\orcidlink{0000-0002-5256-8429}}
\IEEEauthorblockA{
\textit{Technical University of Munich}\\
Heilbronn, Germany \\
stefan.wagner@tum.de}
}

\maketitle

\begin{abstract}
The rapid adoption of AI-powered coding assistants like ChatGPT and other coding-copilots is transforming programming education, raising questions about assessment practices, academic integrity, and skill development. As educators seek alternatives to traditional grading methods susceptible to AI-enabled plagiarism, structured peer assessment could be a promising strategy. This paper presents an empirical study of a rubric-based, anonymized peer review process implemented in a large introductory programming course. 

Students evaluated each other’s final projects (2D game), and their assessments were compared to instructor grades using correlation, mean absolute error, and root mean square error (RMSE). Additionally, reflective surveys from 47 teams captured student perceptions of fairness, grading behavior, and preferences regarding grade aggregation. Results show that peer review can approximate instructor evaluation with moderate accuracy and foster student engagement, evaluative thinking, and interest in providing good feedback to their peers. We discuss these findings for designing scalable, trustworthy peer assessment systems to face the age of AI-assisted coding.
\end{abstract}


\section{Introduction}

AI-powered coding assistants have rapidly emerged as a disruptive force in programming education. Tools like GitHub Copilot and OpenAI’s ChatGPT are now widely accessible – GitHub Copilot became freely available to students upon launch, and ChatGPT reached an unprecedented 100 million users within two months of release \cite{chang2024systematic}. These “\textit{AI coding-copilots}” can automatically generate solutions to various programming tasks and even explain code in plain language. Such capabilities promise to support learners with on-demand help, but they fundamentally challenge traditional teaching and assessment practices in computer science \cite{lau2023ban, wermelinger2023using}. Research indicates that the advent of coding copilots has led to a shift in learning habits, with traditional resources such as YouTube being increasingly supplanted as students' primary source of programming support \cite{berrezueta2024code}. 

Educators and researchers are increasingly concerned about the implications of these tools for academic integrity. Early analyses warn that generative AI systems raise challenges and concerns, particularly academic honesty and plagiarism \cite{simmons2024ai}. If students can obtain AI-generated answers to assignments, it becomes difficult to gauge their accurate understanding or ensure the originality of their work. Instructors have noted that a submission produced with the aid of AI coding-copilots may not accurately reflect a student’s programming ability. There is also a broader worry that ubiquitous AI assistance could normalize new forms of cheating and erode students’ and institutions’ commitment to honesty in coursework \cite{hutson2024rethinking}. As one study put it, academic integrity hangs in the balance with the rise of AI code generators \cite{bin2023use}. 

Beyond integrity issues, the advent of AI coding-copilots has sparked debate about their effect on students’ learning processes and skill development. These tools might give hints, explanations, and examples that scaffold novices' learning. However, many educators fear that over-reliance on AI suggestions will short-circuit the mastery of fundamental programming skills and collaborative work in software projects \cite{akccapinar2024ai}.
A recent survey applied to computer science instructors found that while all were immediately concerned about AI-assisted cheating, they diverged on the long-term pedagogical response. Some advocated strict limitations or “bans” on AI helpers to ensure students continue practicing problem-solving and code writing unaided, reinforcing essential competencies. Others argued for embracing these tools in the curriculum – teaching students with AI and preparing them for a future where AI-assisted coding is the norm \cite{cotton2024chatting}.

The rise of these AI pair-programmers presents a double-edged sword in programming education. Educators are now challenged to rethink course policies and assignment design to uphold educational standards \cite{hutson2024rethinking}. Therefore, this study investigates implementing a structured, anonymized peer review system within a large introductory programming course. Specifically, we analyzed how accurately students evaluate their peers' final projects using a detailed rubric, and how their assessments compare to those of instructors. In addition to quantifying grading reliability through statistical metrics, we gather student reflections on fairness, engagement, and preferences regarding grade aggregation. Drawing on quantitative and qualitative data from 47 teams of three students, this study offers empirical evidence supporting peer assessment as a viable pedagogical strategy for practical programming education in an era increasingly shaped by AI coding tools and reduced opportunities for critical thinking.

\section{Related Work}

Peer review has increasingly been adopted in programming education as a pedagogical tool to engage students in active learning and reflective practice. Several studies report that incorporating peer code review (PCR) activities helps students develop essential programming skills \cite{bradley2019addressing}. Reviewing each other’s code exposes students to diverse solution strategies and helps them learn to give and receive constructive feedback, fostering collaboration and critical thinking. PCR provides students with extensive feedback on their work than is possible from instructors alone, and it trains them in peer evaluation practices. In fact, multiple peer reviews from classmates can yield insights comparable to a single expert review \cite{strickroth2023does}. 

Lin et al. \cite{lin2021using} conducted a randomized controlled trial in a blended introductory programming course and found that students who participated in structured PCR showed significantly improved computational thinking skills and higher engagement levels than a control group. Furthermore, the additional feedback and perspectives gained through peer review can increase students’ time-on-task and encourage deeper reflection. 

However, using peer review in programming education also presents several challenges. One is the reliability and fairness of student-generated feedback, specifically, potential biases or inconsistencies in peer grading.
Novice reviewers may lack the experience to assess code rigorously, leading to variability in the quality of evaluations. Many students are initially skeptical of peer feedback, questioning the credibility of reviews written by colleagues who are still learning the material \cite{alkhalifa2021student}. This skepticism can manifest as resistance to acting on peer comments or reluctance to critique peers openly. 

Another significant challenge is low student engagement in the review process. Studies have reported that participation rates in voluntary peer review activities can be very low (under 20\% of the class) without proper incentives or guidance, and the feedback provided tends to be perfunctory.
For instance, reviewers may leave only brief comments (e.g., one-line remarks), and some of these comments can be partially incorrect or superficial. Reviewers might rush through the task or feel unsure how to constructively critique a peer’s code \cite{brown2019using}. 

Despite the challenges, recent work suggests strategies to mitigate these problems and enhance the effectiveness of peer review. Clear assessment rubrics and training can help standardize evaluations, reducing variability in how different students grade the same code. Anonymizing submissions and reviewers is another tactic to counteract personal bias and encourage honest feedback. Additionally, integrating a reward mechanism or gamification elements has been shown to improve student motivation and the quality of reviews, for example, adding game-like points or badges for thorough feedback led students to write longer, more specific comments in a peer review system \cite{indriasari2023gamification}.

The literature indicates that when peer review is well-structured, with appropriate scaffolding and alignment to learning objectives, it can be a powerful tool in programming education, complementing instructor feedback and actively engaging students in learning \cite{brown2019using, indriasari2020review}. At the same time, authors note that much of the evidence for PCR’s benefits comes from student self-reports, and there is a need for more rigorous studies to objectively measure learning gains and to explore methods for addressing peer feedback reliability.

Bassner et al., \cite{10.1145/3649217.3653543} introduce Iris, an AI-driven virtual tutor integrated into the interactive learning platform Artemis, which provides personalized, context-aware support for computer science students. While Iris focuses on fostering independent problem-solving through subtle hints and Socratic questioning, its design illustrates how AI tools can support the development of critical thinking skills—an essential complement to the peer assessment strategies discussed in this paper.

While prior research has explored the pedagogical benefits and challenges of peer code review in programming education, most studies focus on subjective reports, small-scale implementations, or contexts without comparison to expert evaluation. What differentiates our work is its empirical focus on the accuracy and perception of structured peer assessment in a large, real-world introductory programming course. By combining rubric-based evaluation, statistical alignment with instructor grades, and reflective survey responses from 47 teams, our study offers a comprehensive, data-driven perspective on the reliability, fairness, and pedagogical value of peer review, providing actionable insights for the design of scalable and trustworthy peer evaluation systems in CS education.

\section{Course context}

Fundamentals of Programming (FoP) is an introductory, hands-on course for first-semester students in the Informatics Bachelor's program. No prior programming knowledge is required. The curriculum covers essential topics such as control structures, data types, Object-Oriented Programming (OOP) concepts, streams, graphical user interfaces (GUIs), and recursion \cite{krusche2023introduction}. 
Its assessment is entirely based on practical activities. As shown in Table~\ref{grades}, 60\% of the final grade comes from developing and presenting a final project.

\begin{table}[h]
    \centering
     \caption{Overview of the grading schema for FoP}
    \begin{tabular}{p{5cm}p{2.5cm}}
    \hline
        \textbf{Activity} & \textbf{Grade percentage}  \\ \hline
         Homework exercises & ~20~\%\\ 
         In-class exercises & ~20~\%\\
        \rowcolor{gray!20} \textit{Project development} & ~40~\%\\
        \rowcolor{gray!10}\textit{Project presentation} & ~20~\%\\
        \hline
        \textbf{Total grade} & \textbf{100}~\% \\ \hline
    \end{tabular}
    \label{grades}
\end{table}

The project centers on creating a classic-style 2D game inspired by the concept of \textit{Maze Runner}. The gameplay revolves around steering a character through a complex maze. The main goal is to progress from the starting point to the exit while dealing with hidden traps, hostile entities, and a locked gate that can only be opened with a key. The maze features walls, forming an intricate network of paths, dead ends, and twists. Players must carefully navigate the environment, evade or confront threats on the map, and find keys necessary to unlock the exit and complete the level.

Students had 10 weeks to complete the project, working in teams of three members. Each team can choose an advisor (a course student assistant). While we provide a problem description, a Java project template, a continuous integration (CI) tool, and guidance on using Git for version control, no formal training is given in task coordination, conflict resolution, collaborative project management, or use of artificial intelligence (AI) tools.

Table~\ref{game-rubric} outlines the gameplay, design, and user experience grade distribution. Table~\ref{code-rubric} focuses on software engineering practices such as code and documentation. 
Each category includes several subcomponents totaling a maximum of 100 points (+ up to 10 bonus points). Bonus criteria were included in select categories to reward teams implementing additional features beyond the minimal requirements. 

\begin{table}[h!]
\centering
\caption{Rubric for evaluating gameplay design, graphical quality, sound design, and GUI features of the project.}
\label{game-rubric}
\begin{tabular}{|p{0.78\linewidth}|>{\centering\arraybackslash}p{0.1\linewidth}|}
\hline
\textbf{Category and explanation} & \textbf{Points} \\
\hline
\rowcolor{gray!20}\textbf{Game World}  & \textbf{25} \\
\hline
Entrance and multiple exits & 3 \\
Logical level design, challenges & 7 \\
Reasonable obstacle placement & 4 \\
Enemy AI and interaction & 4 \\
Key/exit mechanism & 2 \\
Pick-up items (power-ups, lives) & 5 \\
\textit{Bonus: Additional features} & +3 \\
\hline
\rowcolor{gray!20}\textbf{Main Character} & \textbf{15} \\
\hline
4-direction movement & 3 \\
Collision detection & 3 \\
Lives lost by triggers (traps) & 3 \\
Camera follows player & 3 \\
Sprinting ability & 3 \\
\textit{Bonus: Player mechanics} & +3 \\
\hline
\rowcolor{gray!20}\textbf{GUI} & \textbf{15} \\
\hline
Main menu options & 4 \\
Persistent HUD (lives, keys, etc.) & 2 \\
Victory/Game Over with return & 4 \\
Scoreboard summary & 5 \\
\textit{Bonus: Responsive/friendly UI} & +1 \\
\hline
\rowcolor{gray!20}\textbf{Sound Design}  & \textbf{10} \\
\hline
Music for gameplay and menus & 6 \\
SFX for actions/events & 4 \\
\textit{Bonus: Layered sounds} & +1 \\
\hline
\rowcolor{gray!20}\textbf{Graphics}  & \textbf{10} \\
\hline
Style matches theme & 3 \\
Resolution scaling & 1 \\
No visual discomfort & 2 \\
No graphical errors/artifacts & 4 \\
\textit{Bonus: Art style/detailing} & +2 \\
\hline
\end{tabular}
\end{table}

\begin{table}[h!]
\centering
\caption{Rubric for assessing code structure and documentation.}
\label{code-rubric}
\begin{tabular}{|p{0.78\linewidth}|>{\centering\arraybackslash}p{0.10\linewidth}|}
\hline
\textbf{Category and explanation} & \textbf{Points} \\
\hline
\rowcolor{gray!20}\textbf{Code Structure} & \textbf{15} \\
\hline
Object-oriented structure & 4 \\
Use of superclasses & 4 \\
No code duplication & 3 \\
Use of delegation/inheritance & 4 \\
\hline
\rowcolor{gray!20}\textbf{Documentation} & \textbf{10} \\
\hline
JavaDoc for methods & 6 \\
Comments in long methods & 2 \\
README with instructions & 2 \\
\hline
\end{tabular}
\end{table}

\section{Methodology}

Once the project's development phase was completed and before the final presentation, we implemented a structured peer-review process.
We developed an algorithm that anonymously assigned each team to review two projects created by other groups in the class, ensuring that no team reviewed its own project or the same project more than once. The reviews were conducted using a detailed grading rubric that outlined the evaluation criteria across several dimensions, including game mechanics, user interface, sound design, graphical quality, code structure, and documentation. This evaluation rubric differentiates from previous related work because it assessed the functional and technical quality of each team’s programming project, not only the code.

\subsection{Assessment process}

\textbf{Gameplay assessment}
Each assessor team must clone and play the game developed by their assigned peers to complete the evaluation rubric detailed in Table \ref{game-rubric}. In addition to the structured criteria, the rubric includes a few provocative questions, such as "Is this game better than yours? Please justify your answer." This prompt encouraged students to evaluate their peers’ work not only against rubric criteria but also against their own project’s quality. Finally, to qualify for the final project presentation, the game must be fully functional, executable, and playable—a determination that is also left to the assessors’ judgment.

\textbf{Code Assessment}
To complete the evaluation described in Table \ref{code-rubric}, students must apply their knowledge of OOP, clean code principles, and library management. This enables them to identify errors and understand their peers' projects to assign fair and accurate grades for each criterion. Additionally, they must assess the quality of the project's documentation, including how clearly the instructions for playing and maintaining the game are written.

Both assessments are submitted through a unified form on a dedicated Confluence page, where only the instructors can access the identities of the assessor and assessed teams. This anonymity ensures impartial evaluations and allows instructors to conduct a thorough analysis afterward.

\subsection{Reliability analysis}

Each project was also evaluated by the course instructors using the same rubric. 
We encouraged all teams to involve every member in discussing each evaluation criterion collaboratively to minimize discrepancies and reduce potential biases that might arise from individual assessments. Furthermore, requiring assessors to justify their assigned grades prompted students to carefully analyze peer submissions and offer fair and constructive feedback.
We compared the peer-assigned scores with those given by instructors to analyze the degree of alignment and identify discrepancies. This comparison served as the basis for evaluating the accuracy of peer assessments. 

We also calculated key statistical metrics for peer evaluations against the instructor's score: Pearson correlation coefficient, mean absolute error (MAE), and root mean square error (RMSE). It provided insights into the potential of integrating peer review as a pedagogical tool in introductory programming education.

\subsection{Reflexive analysis}

To complement the quantitative evaluation of peer grading accuracy, we administered a short reflective survey to all teams immediately after completing their peer review tasks but before receiving the results. The goal was to capture students' perceptions and attitudes toward the peer assessment process. The survey included Likert-scale and open-ended questions. Specifically, teams were asked whether they believed the grades given by peers would be higher or lower than those assigned by instructors. They were asked to self-assess the strictness of their own evaluations using a four-point scale ranging from \textit{Very Strict} to \textit{Not Strict}.

Additionally, teams reflected on whether they considered their evaluations fair and whether they would prefer the average or the highest peer-assigned scores to be counted if discrepancies with instructor grades emerged. Finally, students were asked whether they enjoyed acting as evaluators. This feedback aimed to provide insights into the perceived fairness and acceptability of peer grading and the students' level of engagement in the process.

\section{Results}

\subsection{Reliability analysis results}

Table~\ref{metrics} shows that Peer Review 1 had a moderate positive correlation with instructor grades (\textit{r} = 0.55), a MAE of 9.18, and an RMSE of 14.87. Peer Review 2 exhibited a slightly weaker correlation (\textit{r} = 0.50), with a higher MAE of 10.68 and RMSE of 16.37. 

\begin{table}[h] 
\centering \caption{Comparison of Peer Review Accuracy vs Instructors'} \label{metrics} 
\begin{tabular}{|l|c|c|} 
\hline 
\textbf{Metric} & \textbf{Peer Review 1} & \textbf{Peer Review 2} \\ 
\hline 
Correlation & 0.55 & 0.50 
\\ Mean Absolute Error (MAE) & 9.18 & 10.68 
\\ Root Mean Square Error (RMSE) & 14.87 & 16.36 \\
\hline 
\end{tabular} 
\end{table}

\begin{figure}[h!]
    \centering
    \includegraphics[width=1\linewidth]{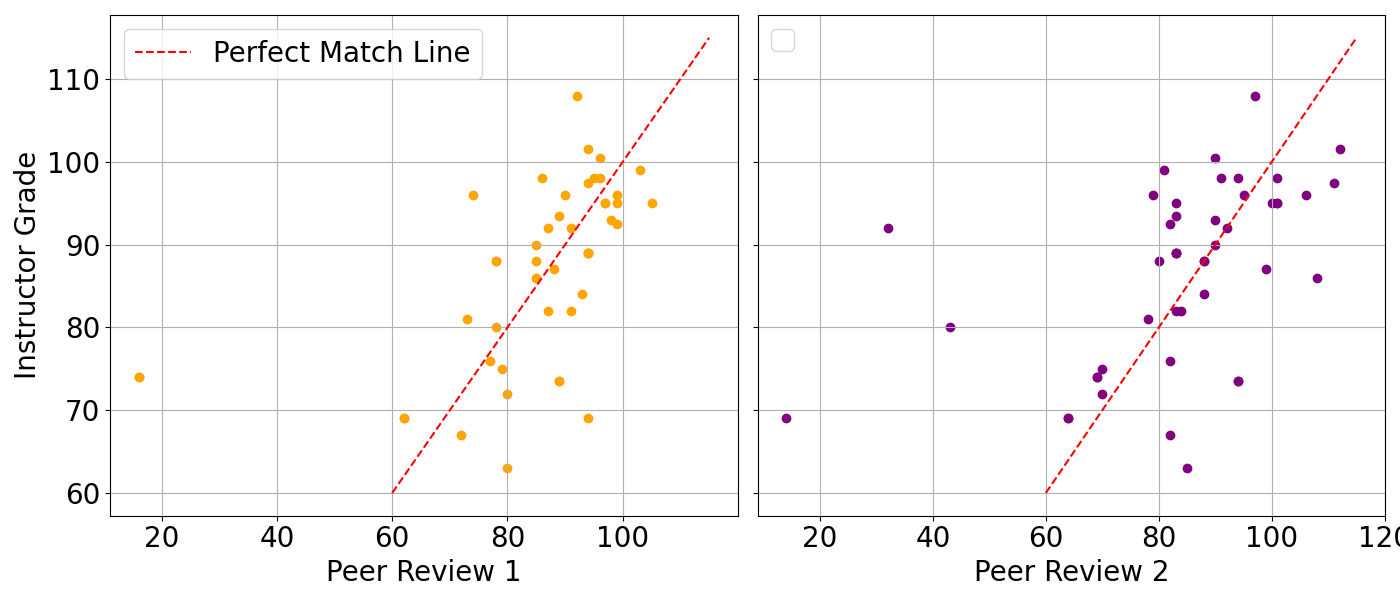}
    \caption{Comparison between peer review scores and instructor grades.}
    \label{Dispersion}
\end{figure}

These results suggest that while peer reviewers were generally aligned with instructors, there were some discrepancies and outliers. This is more evident in Figure~\ref{Dispersion}, where each peer review score is compared against the corresponding instructor grade. While many data points cluster around the Perfect Match Line (the red dashed line) that indicates agreement, several deviations are visible, particularly for Peer Review 2, reinforcing the RMSE results. This means some peer reviewers assigned significantly inflated (to the right of the Perfect Match Line) or deflated grades (to the left of the Perfect Match Line) relative to the instructor’s score.

\begin{figure}[h!]
    \centering
    \includegraphics[width=1\linewidth]{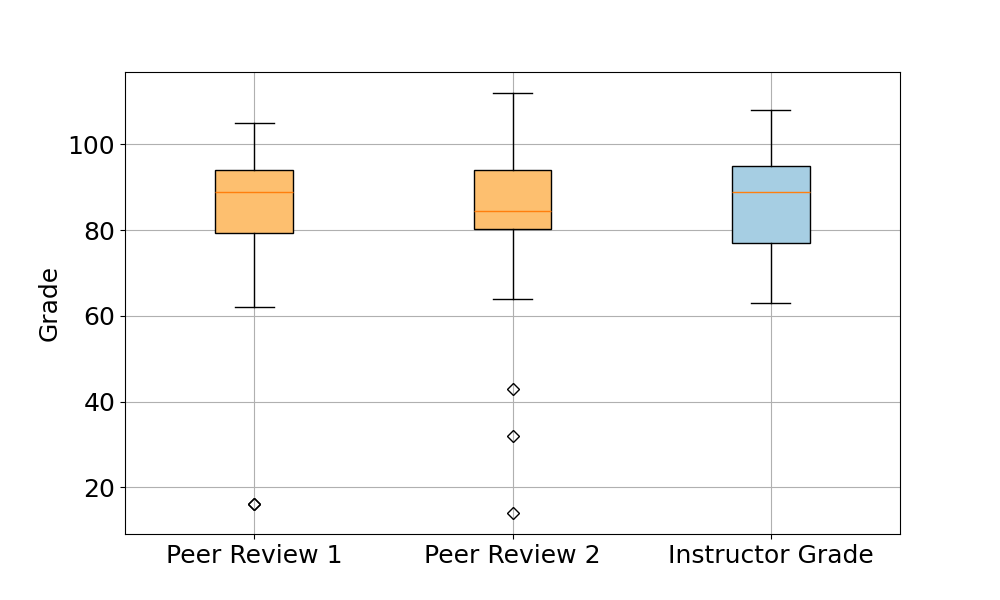}
    \caption{Distribution of Peer Review 1, Peer Review 2, and Instructor Grades highlighting the median, interquartile ranges, and outliers (diamonds).}
    \label{Bloxplot}
\end{figure}

Figure~\ref{Bloxplot} presents that the median and interquartile ranges are relatively close across all three distributions, indicating general consistency in grading. However, several outliers can be observed, particularly in Peer Review 2, indicating that some peer-assigned scores were significantly higher or lower than the instructor’s evaluation.

These results indicate that while peer assessment can often approximate instructor evaluations, variability remains a concern. The peer review 1 was closer to the instructor's grade than the second, suggesting that reviewer training, motivation, or randomness in assignment may influence review quality.

\subsection{Reflexive analysis results}

On the other hand, the results collected in the post-evaluation survey provided insights into student perceptions of fairness, grading behavior, and engagement with the peer review process.

\textbf{Expected Peer vs. Instructor Grades.}
When asked whether they believed their project would receive higher, lower, or similar grades from their peers compared to the instructor, most teams anticipated more favorable peer evaluations. Specifically, 23 teams (49\%) expected a higher grade from their peers, 14 teams (30\%) expected a lower grade, 8 teams (17\%) expected similar outcomes, and 2 teams (4\%) were undecided. This distribution suggests that many students perceive peers as more lenient or empathetic evaluators.

\textbf{Grading Strictness Toward Peers.}
Teams were asked to self-report the level of strictness they applied when evaluating other teams. Of the 47 teams, 26 (55\%) described their grading as “normal,” 15 teams (32\%) as “strict” or “very strict,” and 6 teams (13\%) as mixed or context-dependent. Notably, no team reported being overly lenient, indicating a shared sense of responsibility in applying the evaluation rubric. We corroborated this by Figure \ref{Dispersion}, which shows less dispersed points to the right of the \textit{Perfect Match Line} than to the left.

\textbf{Fairness in Peer Assessment.}
All 47 teams (100\%) believed they provided fair evaluations. Many cited the consistent use of the grading rubric, collaborative discussion within the team, and an effort to remain unbiased despite challenges such as unclear documentation or incomplete features in the reviewed projects. This widespread belief in fairness supports the pedagogical value of structured peer review.

\textbf{Peer Comparison Reflections.}  
In the reflective and comparative question: \textit{“Is this game better than yours? Please justify your answer.”} Interestingly, 28 teams (60\%) answered “No,” stating that their own game was equal to or better than the one they reviewed. Their justifications often included reasons such as greater technical complexity, a more polished user interface, or more complete functionality.
Conversely, 15 teams (32\%) admitted that the reviewed game was better than theirs. These teams frequently cited innovative design choices, a higher level of polish, or additional gameplay features as reasons for their evaluation. The remaining 4 teams (8\%) gave neutral or uncertain answers, often noting that both games had different strengths and weaknesses.
Responses varied from technical comparisons to more subjective impressions, such as creativity and playability. This question proved effective in encouraging students to think critically and reflect on the strengths and limitations of their own work.

Interestingly, when comparing their answers to the final grades, 38 teams (82\%) were accurate in their self-assessment. These results suggest that most students could recognize the relative quality of their projects compared to others, demonstrating a healthy level of self-awareness and evaluative judgment.

\textbf{Preferred Grade Policy.}
To explore student preferences in how peer evaluations should be factored into final grades, teams were asked whether they would prefer the average or the highest peer score in cases of discrepancy. A substantial majority, 32 teams (68\%) preferred the highest score to be used, citing concerns about receiving an unfairly low score from a single reviewer, 12 teams (26\%) preferred the average, and 3 teams (6\%) proposed using the instructor grade alone.

\textbf{Enjoyment of the Evaluator Role.}
We asked about their evaluator experience, 39 teams (83\%) reported enjoying the role. Comments emphasized the opportunity to explore different design ideas, learn from other teams’ solutions, and gain empathy for the grading process. Six teams (13\%) gave neutral responses, while only two teams (4\%) indicated they did not enjoy the task, typically citing incomplete projects as reasons.

\section{Discussion}

Working in teams is a fundamental aspect of learning programming, as it mirrors real-world software development practices and fosters collaborative problem-solving \cite{berrezueta2025assessing}. Teamwork allows students to share diverse perspectives, divide complex tasks, and learn from one another’s strengths and mistakes. This collaborative dynamic becomes even more valuable during peer assessment, where evaluating other teams’ projects as a group encourages critical discussion, more profound understanding of quality code, and more balanced, fair evaluations. 

The findings of this study provide important insights into the feasibility and reliability of using peer assessment in introductory programming education. Our results suggest that structured peer review, supported by detailed rubrics and anonymized processes, can be a reasonably accurate proxy for instructor evaluation. Additionally, we anticipate that a good peer assessment involves the evaluators discussing the rubrics and understanding the project's components before providing feedback and a grade for each one. 

\note{\textbf{Finding 1: Close alignment with instructor grades.} The statistical comparison between peer and instructor evaluations demonstrated a moderate correlation for both review rounds.} The minimal difference, reflected in lower MAE and RMSE values for the Peer Reviewer 1, may be attributed to variations in student engagement, attention to the rubric, or reviewer fatigue. Prior studies have similarly noted that reviewer variability, especially among novice programmers, can affect the consistency of peer evaluations \cite{alkhalifa2021student, brown2019using}. On the other hand, Peer Review 2 exhibited greater dispersion and more outliers, which could stem from the random assignment of weaker or less motivated reviewers.
Despite this variation, the results reaffirm prior literature that peer feedback can approximate expert grading under structured conditions \cite{strickroth2023does, lin2021using}. 

\note{\textbf{Finding 2: Structured rubrics enhance reliability.} The rubric provided a shared framework that likely contributed to the overall alignment with instructor scores.}
Nevertheless, a few projects were significantly underrated by their peers, raising questions about addressing bias and ensuring equity in grading. These variations highlight opportunities and challenges in implementing such systems at scale in more software engineering projects in other courses.

\note{\textbf{Finding 3: Social and cognitive dimensions.} All 47 teams believed their evaluations were fair, even as nearly half anticipated receiving more favorable scores from peers than from instructors.}
This belief in fairness, paired with widespread enjoyment of the evaluator role (83\%) suggests that students took the responsibility seriously and found value in critically reviewing others’ projects. 
These perceptions are consistent with earlier findings that peer review can foster reflection, deepen engagement, and build evaluative judgment among learners \cite{bradley2019addressing}.
However, a few teams preferred to consider only the highest grade in case of discrepancies, highlighting concerns about potential unfairness from a single peer review.

\note{\textbf{Finding 4: Self-awareness and evaluative judgment.} The majority of students were able to accurately assess the relative quality of their own projects compared to those of their peers.}

This reflects a strong sense of self-awareness and evaluative judgment—key competencies in both academic and professional programming contexts. The ability to critically reflect on one's work and compare it fairly against external benchmarks suggests that peer assessment supports grading and fosters metacognitive skills. Such reflective capabilities are especially important in the age of AI-assisted coding, where understanding quality, originality, and design trade-offs becomes as important as writing code itself.

While statistical alignment with instructor scores is a positive indicator, it is equally important to cultivate transparency and offer feedback loops that allow learners to reflect on the quality of their evaluations. In future iterations, additional training, calibration exercises, or use of tutor-assisted scoring could further improve the consistency and fairness of peer assessments.

\section{Conclusions}

This study examined the effectiveness of structured, anonymized peer assessment in a large-scale introductory programming course, demonstrating that students can reliably evaluate each other’s work using detailed rubrics. The alignment between peer and instructor evaluations and positive student perceptions of fairness and engagement underscores the pedagogical potential of peer review in programming education.
Importantly, peer assessment also promotes key competencies that are increasingly vital in the age of AI copilots—namely, critical thinking, evaluative judgment, and collaborative reflection. As coding assistants automate more aspects of programming tasks, assessing, critiquing, and understanding quality code becomes essential for learners to remain active participants in the development process, rather than passive users of AI-generated solutions.

As programming education evolves in response to technological shifts, scalable and participatory assessment models will become increasingly important. Future research should explore ways to refine peer review systems through reviewer training, calibration sessions, or adaptive weighting of scores. Additionally, integrating formative peer assessment across multiple stages of the project lifecycle, not only at the end, could offer ongoing feedback and learning opportunities. Investigating the long-term impact of such practices on learning outcomes, ethical reasoning, and professional readiness in AI-augmented environments remains a valuable direction for further work.

Ultimately, our findings suggest that well-designed peer assessment is a practical response to the challenges posed by AI coding tools and a meaningful strategy for empowering students in reflective, responsible, and collaborative learning environments.

\balance
\bibliographystyle{ieeetr}
\bibliography{From_Coders_to_Critics}

\end{document}